\documentclass[%
 aip,
 jmp,%
 amsmath,amssymb,
 reprint,%
]{revtex4-2}

\usepackage{graphicx}
\usepackage{dcolumn}
\usepackage{bm}
\usepackage{cleveref}

\begin{document}

\preprint{APS/123-QED}
\title[Quantum theory of asymptotically flat spacetimes]{Quantum theory of asymptotically flat spacetimes}

\author{Carlos N. Kozameh}
\email{carlos.kozameh@unc.edu.ar}
\author{Lautaro Zapata-Altuna}%
\affiliation{FaMAF, Universidad Nacional de C\'ordoba, 5000, C\'ordoba, Argentina}%

\date{\today}

\begin{abstract}

A quantum theory of asymptotically flat space-times is presented using the solutions of the Null Surface Formulation (NSF) field equations in a perturbative scheme. Free-field commutation relations are given for the null free data of NSF at future or past null infinity, and the main variables of NSF are then used to define the interior points of the space-time as labels and the metric of the space-time as a derived quantum operator. A non-trivial scattering of gravitons is given.
\end{abstract}

\maketitle

\section{\label{int} Introduction}
In this work we present a quantum theory for a certain class of asymptotically flat space-times that are vacuum, globally hyperbolic, and contain no horizons\cite{dominguez1997phase}. These spacetimes are called classical gravitons and represent the self-interaction of incoming gravitational radiation at past null infinity that produces outgoing radiation at future null infinity. 
Although there are no explicit nontrivial solutions of vacuum relativity satisfying the requirements of a classical graviton spacetime, one could use the theorems on global stability of Minkowski space by Chrusciel-Delay\cite{Chrusciel2002} , to  argue that  there is a nontrivial neighborhood of Minkowski space (in the solution space) with this property. Thus, the results presented in this work apply for this subclass of classical gravitons.

The approach uses the Null Surface Formulation of General Relativity, or NSF for brevity, which gives an alternative formulation of an asymptotically flat space-time\cite{bordcoch2016asymptotic}. Instead of a bounded manifold with a regular metric, the NSF introduces a real function $Z$ with a dual meaning. 
\begin{itemize}
    \item Given an asymptotically flat space-time with coordinates $x^a$ and future null infinity with Bondi coordinates $(u,\zeta, \bar{\zeta})$, the equation

\begin{equation}\label{Z_1}
    u=Z(x^a,\zeta, \bar{\zeta}),
\end{equation}
represents the intersection of the future light cone from $x^a$ with future null infinity and it is called a null cone cut of null infinity or simply a null cut. In general, this function is locally defined since caustics and self-intersections develop on the future light cone of a point. 

    \item For a fixed value of $(u,\zeta, \bar{\zeta})$, the hypersurface $Z(x^a,\zeta, \bar{\zeta})=u= const$ is the past the null cone from the point $(u,\zeta, \bar{\zeta})$ at future null infinity. As we will see below, $Z$ is defined in the sphere bundle of null directions over spacetime and it is a fundamental variable in the NSF.
\end{itemize}

NSF also introduces a conformal factor $\Omega$ on the spacetime that is used to relate the conformal structure given by $Z$ with the space-time metric. 

Since this work aims to to give a global construction, namely, the non-trivial scattering of gravitons, we must assume that null cuts are regular functions at both future and past null infinity. 
This is a very important and non-trivial restriction on the subclass of classical graviton spacetimes and one should check whether or not there are solutions which lie in a small non-linear neighborhood of Minkowski space  (in the space of asymptotically flat solutions) for which the null cuts are regular closed surfaces at null infinity. Since this assumption is true in Minkowski space (where $Z$ is a regular function of $l=0,1$ spherical harmonics), one can use the stability theorems of Minkowski space\nocite{Chrusciel2002} to argue that there is a small nonlinear neighborhood of Minkowski space  for which $Z(x^a,\zeta,\bar\zeta)$ remains a regular function at null infinity.

It is worth mentioning that a linearized version of NSF has been quantized\cite{frittelli1997fuzzy,frittelli1997quantization}. In these works the points of the underlying spacetime arise as the constants of integration of the linearized field equations for $Z$ and the quantum $Z$ yields fuzzy null surfaces on the solution space. In this work the points of the space-time will also be labels associated with the constant of integration of the field equations.
Finally, in the classical NSF the free null data correspond to the incoming or outgoing gravitational waves. The quantum version would be incoming or outgoing free gravitons, the quantum particles of gravity.

In summarizing, in NSF the only stage given ab initio is the future and past null infinity together with the free data, the News tensor or the Bondi shear on a Bondi coordinate system.

The solution space of the field equations yields the points of the space-time for which the level surfaces of $Z$ are null. It is worth mentioning that in the NSF the standard Ricci flat equation is complemented with two extra equations that are called metricity conditions. They are kinematical equations in the sense that any Lorentzian metric must satisfy those equations. The Ricci flat equation determines the dynamics of  the classical graviton. 

To obtain a quantum theory of NSF we use a very important mathematical structure given by the commutation relations of the creation-annihilation operators at either future or past null infinity. This has been provided by A.Ashtekar in the so-called asymptotic quantization of the gravitational field\cite{ashtekar1981asymptotic,ashtekar1987asymptotic}. In his setup the stage is the future null infinity of any asymptotically flat space-time.  This once and for all construction of the null boundary represents the stage, and the null data given by the News tensor is the free field that is transformed to a quantum operator with commutation relations. A similar construction could be given at past null infinity and those quantum fields given at either boundary could be interpreted as the outgoing or incoming gravitons of the quantum theory. Since these null data are the only source given on the NSF equations, one obtains a quantum NSF.

It is worth mentioning that there are no constraints in the theory since the null data are given freely on the initial surface. Thus, in principle, one has at our disposal a quantum theory of gravitons and the idea is to test the theory to see whether or not it gives sensible results.

At a classical level $Z$ and $\Omega$ functionally depend on the News tensor given on the boundary. Thus, a quantized null data automatically yield quantum operators via the field equations. In this sense the quantized operators are auxiliary fields that are used to define the scattering of quantum gravitons.

In this work, we will solve perturbately the NSF equations to second order and provide a quantization procedure for the perturbed solution obtaining a quantum operator for the metric, null surfaces, etc. Moreover, adding an extra assumption on the nature of the null boundary (which remains classical), we obtain the scattering of quantum gravitons. There is a caveat in this procedure, namely, for this construction to be regular at a classical level, the null data must be small, i.e. one adds an adimensional parameter $\epsilon$ to the Bondi data and assumes that it is small (for $\epsilon=0$ one obtains flat space). Using the Sachs equations for the shear and divergence of null cones in the unphysical metric, one can show that if $\epsilon$ is sufficiently large, singularities develop before reaching null infinity and the null cuts fail to be regular closed 2-surfaces. At a classical level this sets a limit on the intensity of the incoming or outgoing gravitational radiation. At a quantum level this excludes coherent states that are peaked at large initial data. These coherent quantum states cannot be included in the present approach.

 The outline of this approach is the following: starting with NSF and its main variables, one then obtains the metric of a classical graviton spacetime\cite{bordcoch2016asymptotic}. We also use the NSF equations for the main variables of the formalism to second order. Depending on whether the free data at future or past null infinities is selected, the NSF solutions are called advanced or retarded and labeled with a $+$ or $-$ sign. Imposing the condition that the advanced and retarded solutions of the NSF yield the same metric we obtain the non-trivial scattering of incoming and outgoing gravitational waves at a classical level\cite{bordcoch2023asymptotic}. Finally, we promote the News tensor at future or past null infinity to a quantum operator with commutation relations. This yields a relationship between the incoming and outgoing creation-annihilation operators and obtains the non-trivial terms of the scattering process. 

The are in the literature other perturbative approaches\cite{Weinberg1979,Parker-Toms2009,strominger2018lectures} to obtain an S matrix for quantum gravity. In particular, there are some results\cite{strominger2020} that should be related to this work. It is worth exploring the similarities and differences with this work.

The paper is organized as follows. In \ref{sectionII} we give a brief review of the NSF giving a kinematical description of the main variables and the field equations that are valid up to the second order in a perturbation procedure. We also give the relation between the metric and the NSF variables for the linear case and show that the map between the past and future null radiation data is the identity. In \ref{sectionIII} we review some results that arise in the asymptotic quantization of the gravitational field that are useful for developing the main results of this work. In \ref{sectionIV} we derive advanced and retarded quantum solutions to the second order. They are then used to obtain the scattering of gravitons showing a non-trivial interaction that can be explicitly displayed. Finally, we finish the work with a summary and conclusions in \ref{sectionV}.

\section{\label{sectionII}The Null Surface Formulation}

In this section we review the main results of NSF and compute the second-order perturbation solution that is then used to quantize the theory.
The Null Surface Formulation recasts general relativity as a theory of null surfaces with field equations that are equivalent to the Einstein equations\cite{kozameh1983}. In this work we will be using the second-order approximation of the field equations given in 2016\cite{bordcoch2016asymptotic}.

The review of NSF presented in this section is suited to the class of space-times we are dealing with in this work. We will use future null infinity to define  one of the main variables but the construction also applies to past null infinity.

 Given a classical graviton space-time and future null infinity with Bondi coordinates $(u,\zeta,\bar{\zeta})$ we define $Z$ as the intersection of the future null cone from the point $x^a$ with the null boundary. This intersection can be written in Bondi coordinates as $u=Z(x^a,\zeta,\bar{\zeta})$. In principle, this is a local construction since a null cone in a non-flat geometry develops caustics and self-intersections, but assuming the classical graviton spacetime is a small deviation of Minkowski space\cite{Chrusciel2002}, one can assume the function $Z$ is regular and differentiable, Thus, the null cut is a closed 2-surface at both future and past null infinity.
The scalar $Z$ has a second meaning which is more subtle and is a consequence of the reciprocity theorem for null geodesic congruences. Given the function $Z(x^a,\zeta,\bar{\zeta})$ previously obtained, and keeping the coordinates $(u,\zeta,\bar{\zeta})$ fixed, the level surface of $Z$, that is, $Z(x^a,\zeta,\bar{\zeta})=u= const$ is the past null cone from the point $(u,\zeta,\bar{\zeta})$ at null infinity. From the point of view of the physical spacetime, this surface starts as a null plane at future null infinity and starts developing shear and divergence as it goes inside the space-time. In Minkowski space it is a flat null plane and it is suitably adapted to define a quantum particle. This fact will be used later.

Since $Z= const.$ are null hypersurfaces, they satisfy
\begin{equation}\label{g00}
g^{ab}(x^a)\partial_a Z(x^a,\zeta,\bar{\zeta})\partial_b Z(x^a,\zeta,\bar{\zeta})=0,
\end{equation}
for the underlying metric of space-time. Note that the equation is satisfied for every value of $(\zeta,\bar{\zeta})$.  Thus, the scalar $Z$ is defined on the six-dimensional bundle of null directions over space-time. It is possible to write the conformal metric in terms of the null cuts by first defining a $S^2$ family of coordinate systems, one for each value of $(\zeta, \bar\zeta)$, constructed from knowledge of $Z$, using the gradient basis of that coordinate system and then writing the non-trivial components of the conformal metrics in terms of $Z$. Only a conformal metric can be obtained by this procedure since one can multiply eq. (\ref{g00} by any function of $x^a$ and is also satisfied by the same $Z$.

The $(\zeta, \bar\zeta)$ coordinate system that is constructed from $Z$ and its derivatives on the sphere, is given by

$$\theta^i(x^a, \zeta, \bar\zeta)= (u, w, \bar{w},r)=(Z, \eth Z, \bar\eth Z, \eth \bar\eth Z).$$

Note that the four scalars that define the coordinates $\theta^i$ are, by assumption, smooth functions of $x^a$ since the null cuts are regular 2-surfaces for every point of the space-time. From the point of view of the null cuts, all the points $x^a$ whose null cuts reach null infinity at $(u, \zeta, \bar\zeta)$ have the same value of $Z$. Moreover, all the points that lie on a null geodesic $x^a (s)$,with $s$ an affine length, will also have the same value of $(\eth Z, \bar\eth Z)$, and each value of $s$ gives a corresponding value of $\eth \bar\eth Z$.

Conversely, for a fixed value of $(\zeta, \bar\zeta)$, $Z=u=const.$ is the past null cone from the point $(u, \zeta, \bar\zeta)$ at null infinity. On that null surface $(w, \bar{w})$ identifies a null geodesic and $r$ a point on the geodesic.

The metric reconstruction algorithm must take a sufficient number of $\eth$ and $\bar\eth$ derivatives in the eq. $(\ref{g00})$ so that all the components of the conformal metric are obtained. By an explicit calculation, one finds that the nontrivial components are explicit functions of $\Lambda(x^a, \zeta, \bar\zeta)=\eth ^2 Z$ while the tenth component, denoted by $\Omega^2$ remains undetermined. The equation reads

\begin{equation}\label{conformal}
g^{ab}(x^a)=\Omega^2 h^{ab}[\Lambda].
\end{equation}
where $\Omega^2$ is defined as 
 \begin{equation}\label{g01}
\Omega^2= g^{ab}\partial_aZ\partial_b\eth \bar\eth Z=\frac{\partial r}{\partial s}
\end{equation}
with $s$ an affine length of the null geodesic on the past null cone. 

It is worth mentioning that both $\frac{\partial}{\partial s}^a=g^{ab}\partial_bZ $  and $\frac{\partial}{\partial r}^a$ are null vectors that are related via 

$$ \frac{\partial}{\partial s}^a=\Omega^2\frac{\partial}{\partial r}^a$$
and by the asymptotic conditions $r$ is an injective function of the affine length $s$. One can show that on a generic asymptotically flat space time $r$  and $s$ go to infinity at null infinity but $r$ blows up at a caustic whereas $s$ remains finite at that point$\cite{bordcoch2016asymptotic}$. However, in this particular construction where the spacetime is a neighborhood of of Minkowski space, one is assuming that $Z$ is a regular function and via the reciprocity theorem $r$ is a regular function at every point $x^a$. It follows that the global requirement of regular cuts implies a coordinate system that is globally given on a spacetime that lies in a small non-linear neighborhood of Minkowski space (in the space of asymptotically flat
solutions). For more details, see $\cite{bordcoch2016asymptotic}$.

Although the $x^a$ dependence of $\Omega$ is arbitrary at this stage, its $(\zeta, \bar\zeta)$ dependence is not since $g^{ab}$ does not depend on $(\zeta, \bar\zeta)$ whereas $h^{ab}$ does. One can show that directly from
$$
    \eth^2 \bar{\eth}^2(g^{ab}\partial_a Z\partial_b Z)=0,
$$
one obtains a relationship between $\Omega$ and $\Lambda$, namely,
\begin{equation}\label{g02}
\eth \bar{\eth}(\Omega^2)=\Omega^2(\frac{\partial (\bar{\eth}^2 \Lambda)}{\partial 
r} -h^{ab}\partial_a \Lambda\partial_b \bar{\Lambda}).
\end{equation}
This equation is called the first metricity condition. It is easy to see that the above equation is invariant under rescaling of $g^{ab}$ i.e. under multiplication of $\Omega$ by an arbitrary function of $x^a$.

Since we use $Z$ as our main variable, it is important—at a kinematical level—to ask whether the metric construction procedure given above can be applied to an arbitrary function $Z(x^a, \zeta, \bar{\zeta})$. The answer clearly is no, $Z$ cannot be arbitrary because the conformal metric is determined by 9 functions, whereas we must satisfy an infinite number of equations (one for each value of $(\zeta, \bar{\zeta})$). Thus, a kinematic condition must be imposed on $Z$ to ensure that its level surfaces are null with respect to the conformal metric obtained from the construction procedure. This kinematic condition is given by:
  
  $$
    \eth^3 (g^{ab}\partial_a Z\partial_b Z)=0 \implies 
    h^{ab}(3 \partial_a \eth Z\partial_b \Lambda + \partial_a Z \partial_b \eth \Lambda)=0.
$$
and it  can be rewritten as 
\begin{equation}\label{Wunchsmann}
\frac{\partial \eth \Lambda}{\partial r}+3 h^{wi}\partial_i \Lambda=0,
\end{equation}
This equation is called the second metricity condition. Only for functions $\Lambda$ that satisfy this equation it is possible to show that $Z=const.$ is a null surface. To fully understand the meaning of this equation, we go back to its definition. Inserting the inverse relationship between $x^a$ and $\theta^i$ we rewrite the equation as
\begin{equation}\label{Lambda}
\eth^2 Z = \Lambda(Z, \eth Z, \bar\eth Z, \eth \bar\eth Z,\zeta,\bar{\zeta}),
\end{equation}
which is a PDE on the sphere for Z for a given function Lambda that depends on this six variables. Note that the metricity condition (\ref{Wunchsmann}) is a PDE for $\Lambda$ in terms of the natural coordinates of equation (\ref{Lambda}). Only for $\Lambda$s that satisfy the metricity condition can a conformal metric be obtained in the solution space of eq. (\ref{Lambda}). The solution space is given by the kernel of the $\eth^2$ operator and are 4 numbers that we can label as $x^a$, i.e. the solution space of the above equation becomes the space-time points.

It is remarkable that the metricity condition for the three-dimensional version of the NSF coincides with the so-called Wünschmann condition \cite{Wunschmann1905} which Cartan derived while studying the equivalence classes of ODEs under diffeomorphisms \cite{Cartan1938,Cartan1941,Chern1940},. 
For systems satisfying the Wünschmann condition in three dimensions, one can construct a conformal metric, and diffeomorphisms between such metrics on the solution space yield equivalent classes of diffeomorphic ODEs. 

The field equations for NSF take a simple form in terms of this coordinate system. Directly from the relationship between $g_{ab}$ and $h_{ab}$, the trace free Ricci flat equations read \cite{bordcoch2016asymptotic,wald2010general}:
\begin{align}\label{EinsteinNSF}
    2\partial _{r}^{2}\Omega  = R_{rr}[h] \Omega,
\end{align}
with the component $R_{rr}$ given by

\begin{eqnarray}
R_{rr}[h]&=&\frac{1}{4q}\partial _{r}^{2}\Lambda \partial _{r}^{2}\bar{\Lambda}+\frac{%
3}{8q^{2}}(\partial _{r}q)^{2}-\frac{1}{4q}\partial _{r}^{2}q,
\label{R11}
\end{eqnarray}
$$q =1-\partial _{r}\Lambda \partial _{r}\bar{\Lambda},$$
where we have adopted the notation $\partial _{r}=\frac{\partial}{\partial r}$ for simplicity. It is clear from the above equations that $R_{ab}[h]$ vanishes when $\Lambda=0$. 

The three scalar equations (\ref{g01}), (\ref{Wunchsmann}),  and (\ref{EinsteinNSF}) are completely equivalent to the vacuum Einstein equations for a metric $g_{ab}$. If we integrate (\ref{g01}) and (\ref{Lambda}) along a null geodesic and ask for regularity conditions on the integrals of (\ref{Lambda}) using the peeling theorem, then we get the vacuum, asymptotically flat solutions, the so-called classical gravitons. They represent the non-linear interaction of incoming gravitational waves that are freely given at future null infinity. This backward time direction can be thought of as an advanced solution of a generalized wave equation. One can also obtain the retarded graviton solutions where outgoing gravitational waves are freely given at past null infinity and evolve to the future giving a Ricci flat metric. The final form of the classical graviton field equations is given by \cite{bordcoch2016asymptotic},
\begin{equation}\label{Lambda2}
\bar{\eth}^2 \eth^2 Z= \eth^2 \bar{\sigma}(Z,\zeta,\bar{\zeta})+ \bar{\eth}^2 \sigma(Z,\zeta,\bar{\zeta})+\Sigma^+(Z,\zeta,\bar{\zeta}) -\int _r^\infty(\Omega^{-2}\eth \bar{\eth}(\Omega^2)+h^{ab}\partial_a \Lambda\partial_b \bar{\Lambda})dr',
\end{equation}
with
$\Sigma^+(Z,\zeta,\bar{\zeta})=\int _{-\infty}^Z \dot{\sigma}\bar{\dot{\sigma}} du$, 
and

\begin{equation}\label{Omega'}
\Omega= 1 +\int _r^\infty dr'\int _{r'}^\infty R_{rr}[h] \Omega dr'',
\end{equation}

In the above equations, $\sigma(u,\zeta,\bar{\zeta})$ is the Bondi shear given at future null infinity, $\dot{\sigma}$ is the derivative with respect to the Bondi time $u$ and $\Sigma^+(Z,\zeta,\bar{\zeta})$ is the change in the mass aspect. Integrating $\Sigma^+(Z,\zeta,\bar{\zeta})$ on the cut yields the Bondi mass change due to loss of gravitational radiation. The complex shear represents the two degrees of freedom of the gravitational field, and the solutions of the above equations are functionally dependent on $\sigma(u,\zeta,\bar{\zeta})$.

Equation (\ref{Omega'}) together with conditions (\ref{Wunchsmann}) and (\ref{Lambda2}) are necessary and sufficient to construct a classical graviton spacetime. In this approach, the function
$\Lambda$ plays an important role since $h^{ab}$ is completely given in terms of this function, and its vanishing yields a flat metric. Thus, one can implement a perturbation procedure directly from knowledge of $\Lambda$ and write down the lowest non-trivial formulation from a linearized approximation.

Since the Bondi shear plays the role of a source term in the NSF equations it is worthwhile to make a few remarks on the regularity behaviour of the null cone cuts.

By assumption, the spacetime is a Ricci flat, asymptotically flat spacetime without singularities that represent incoming or outgoing gravitational radiation, i.e., what we define as a classical graviton (see reference \cite{dominguez1997phase}). Those space-times are constructed from the free Bondi data given at null infinity contained in the News function, which is assumed to have compact support. If the News function vanishes the only solution is Minkowski space, and one can add a parameter $\epsilon$ to represent how much it deviates from flat space. Assuming the graviton spacetime is a small deviation of flat space\cite{Chrusciel2002}, one can solve the optical equations using the unphysical metric and since the geometry is regular everywhere, the unphysical affine length reaches either past or future null infinity at a finite distance, it is possible to show that for a sufficiently small value of the parameter $\epsilon$ the divergence of the null cone from a point does not blow up at the boundary and its intersection with null infinity is a closed 2-surface.  Assuming the null cone cuts are closed regular 2-surfaces, one can then introduce a perturbative approach to solve field equations. The perturbed solutions should also be non-singular otherwise the perturbation procedure breaks down. We give below the solutions up to second order and we absorbe the parameter $\epsilon$ in the News function.

\subsection{The first order solution}

The linearized version of the last two equations is obtained by first giving the zeroth-order solution that yields a flat metric, namely,

\begin{align} \label{Z0}
Z_0 = x^a l_a,  \; \; \Omega_0= 1,
\end{align}
with $l^a$ a null vector defined as $l^a=\frac{1}{\sqrt{2}}(1,\hat{r}^i)$ and $\hat{r}^i$ the unit vector on the sphere of null directions, and $x^a$ a point in the flat spacetime.\\
One then writes down a linearized departure from the zeroth order solution as
\begin{align}
Z = x^a l_a + Z_1, \; \; \Omega = 1+\Omega_1.
\end{align}

The equation of motion for $\Omega_1$,
\begin{align}\label{EinsteinNSF2}
    2\frac{\partial^2\Omega_1}{\partial s^2}  = 0,
\end{align}
yields a trivial solution when regularity conditions are imposed while the equation for $Z_1$ is given by,

\begin{align}\label{Z1}
&\bar{\eth}^2\eth^2 Z_1 =\bar{\eth}^2 \sigma(Z_0,\zeta,\bar{\zeta}) + \eth^2 \bar{\sigma}(Z_0,\zeta,\bar{\zeta}) +\mathcal{O}(\Lambda^2).
\end{align}

 Eq. (\ref{Z1}) is a non-homogeneous 4th order eliptic equation on the sphere and its solution can be found by convoluting the inhomogeneity with the corresponding Green function \cite{ivancovich1989green}. 
 The solution reads,

 \begin{equation}\label{Z_1b}
    Z^{+}_{1}(x^{a},\zeta)=\oint_{S^{2}}G_{00'}(\zeta,\zeta')\left(\eth'^{2}\overline{\sigma}^{+}(x^{a}l'^{+}_{a},\zeta')+\overline{\eth'^{2}}\sigma^{+}(x^{a}l'^{+}_{a},\zeta')\right)dS'
\end{equation}

 with 

\begin{equation}\label{G_00'}
    G_{00'}(\zeta,\zeta')=\dfrac{1}{4\pi}l^{a}l'_{a}ln(l^{a}l'_{a}),
\end{equation}
and $l^{a}$ given by,
\begin{equation}
    l^{a} = \frac{1}{\sqrt{2}(1 + \zeta\overline{\zeta})}\left(1 + \zeta\overline{\zeta}, \zeta + \overline{\zeta}, -i(\zeta - \overline{\zeta}), -1 + \zeta\overline{\zeta}\right).\label{5555-1} \\
\end{equation}

 In the above equation we have simplified the notation and omitted $\bar{\zeta}$. We will keep this change on notation for the remainder of this work but there is no implication of complex analytic functions since all our variables are real and/or given on the complex stereographic functions on the real sphere. 
 
 It is worth  pointing out that (\ref{Z1}) is given at future null infinity and the points $x^a$ in the equation are simply constants of integration. The connection with the underlying spacetime is given when $Z_0$ is interpreted as the null cone cut for a flat spacetime and for that one presupposes the existence of this null boundary. The same happens when giving a similar construction for past null infinity. Moreover, the two null boundaries are connected via spacelike infinity which is a point for a flat spacetime.
 
\subsection{The second order solution}

The second order $Z_2$, $\Omega_2$ satisfy the following equations,

\begin{equation}\label{Z_2}
\bar{\eth}^2 \eth^2 Z_2= \eth^2 \bar{\sigma}(Z_1,\zeta)+ \bar{\eth}^2 \sigma(Z_1,\zeta)+\Sigma(Z_0,\zeta) -2\int _{r}^\infty(\eth \bar{\eth}(\Omega_2)+\eta^{ab}\partial_a \Lambda_1\partial_b \bar{\Lambda_1})dr,
\end{equation}
\begin{align}\label{Omega2}
    \partial _{r}^{2}(8\Omega_2-\partial _{r}\Lambda_1 \partial _{r}\bar{\Lambda}_1)= \partial _{r}^{2}\Lambda_1 \partial _{r}^{2}\bar{\Lambda}_1.
\end{align}

Equation (\ref{Z_2})can be integrated with the same Green function (\ref{G_00'}) whereas the solution to the ODE (\ref{Omega2}) is given by 
\begin{align}\label{Omega2sol}
    8\Omega_2=\partial _{r}\Lambda_1 \partial _{r}\bar{\Lambda}_1+ \int _{r}^\infty dr' \int _{r'}^\infty dr''\partial _{r''}^{2}\Lambda_1 \partial _{r''}^{2}\bar{\Lambda}_1.
\end{align}
with $\Lambda_{1}=\eth^{2}Z_{1}$. 

\subsection{Higher order solutions}
As one can see, eqs. (\ref{Wunchsmann}) and (\ref{g01}) are polynomial in powers of $\Lambda$ whereas (\ref{EinsteinNSF}) has fractional terms with $q$ in the denominator. Multiplying eq. (\ref{EinsteinNSF}) by $q^2$ one gets a polynomial equation that can be solved perturbately assuming that $\Lambda$ is small. We will return to this issue when the cut $Z$ is quantized.

\subsection{The metric tensor and null surfaces}

As it was mentioned before, $\partial_{a}Z$ is a null covector and satisfies,
\begin{equation}\nonumber
g^{ab}\partial_{a}Z\partial_{b}Z=0.
\end{equation}

It also follows from the field equations that $Z$ has a functional dependence on the free null data $\sigma$. Thus, assuming the free data is small, one can write down a  perturbation series for \ref{g00} relating $g^{ab}$ with the perturbed solutions of $Z$. We thus write
\begin{equation}
\sum_{n=0}^{\infty}\sum_{r+s=0}^{n}g_{n-r-s}^{ab}\partial_{a}Z_{r}\partial
_{b}Z_{s}=0\label{gn},
\end{equation}
where
\begin{equation}
\sum_{n=0}^{\infty}g_{n}^{ab}=g_{0}^{ab}+g_{1}^{ab}+g_{2}^{ab}+...=(1+\Omega_1+\Omega_2+...)^2(\eta^{ab}+h_{1}^{ab}+h_{2}^{ab}+...)
\end{equation}
with $\eta^{ab}$
the flat metric and the labels $_{1},_{2},...$ the different orders of the NSF variables.
\begin{itemize}
\item Taking $n=0$ in (\ref{gn}) we have:
\begin{equation}
\eta^{ab}\partial_{a}Z_{0}\partial_{b}Z_{0}=0\label{g0}.
\end{equation}
Taking $\partial_{a}$ on \cref{Z0} we obtain $\partial
_{a}Z_{0}=l_{a}$. then the expression(\ref{g0}) can be written as
\begin{equation}
\eta^{ab} l_{a} l_{b}=0\label{eta}.
\end{equation}
Taking $\eth$ and $\bar{\eth}$ on (\ref{eta}) one gets all the metric components in the flat null tetrad \cite{newman2005tensorial}

\item Taking $n=1$ in (\ref{gn}) one gets,
\begin{equation}\label{h1}
h_{1}^{ab}l_{a}l_{b}+2\eta^{ab}l_{a}\partial_{b}Z_{1}=0,
\end{equation}
since $\Omega_1$ vanishes at the linearized approximation.
The above expression can be rewritten as
\begin{equation}
h_{1ab}l^{a}l^{b}+2l^{a}\partial_{a}Z_{1}=0,\nonumber
\end{equation}
from which one can obtain all the components of $h_{1ab}$. Note that $l^{a}$ is not a null vector of $h_{1ab}$ but nevertheless it is useful to determine all the components of the linearized metric. Following a straightforward calculation outlined in \cite{bordcoch2023asymptotic}, one obtains,
\begin{equation}\label{métricaprimerorden}
 h_{1ab}(x)=\dfrac{-1}{2\pi}\oint_{S^{2}}\left(m'_{a}m'_{b}\dot{\overline{\sigma}}^{+}(x^{a}l'^{+}_{a},\zeta')+\overline{m}'_{a}\overline{m}'_{b}\dot{\sigma}^{+}(x^{a}l'^{+}_{a},\zeta')\right)dS'.
\end{equation}
We see that $h_{1ab}(x)$ has a linear dependence on the free data at null infinity.

\item Taking $n=2$ on (\ref{gn}), we get the second order term,
\begin{equation}\label{h2}
h_{2ab}l^{a}l^{b}+2h_{1}^{ab}l_{a}\partial_{b}Z_{1}+ 2 l^{a}\partial_{a}Z_{2}=0,
\end{equation}
from which $h_{2ab}$ can be obtained by repeated $\eth$ and $\bar{\eth}$ operations on (\ref{h2}).

Up to second order, the metric of the spacetime can be written as
\begin{equation}
g_{ab}=\eta_{ab}+h_{1ab}+2\Omega_2 \eta_{ab}+h_{2ab},\label{g2}%
\end{equation}
\end{itemize}
where $\Omega_2$ and $Z_2$ are obtained from the second order field equations and $h_{2ab}$ is algebraically related to $Z_2$ via \cref{h2}.

The above construction can be done with free data given at future or past null infinity. In each case the solution is labelled as $Z^+$ or $Z^-$ respectively and it is analogous to the advanced or retarded solutions that can be constructed for the solutions of the wave equation. Following the algebraic relationship between the pair $(Z,\Omega)$ and the metric of the spactime one can construct an advanced, $g^+_{ab}$, or retarded, $g^+_{ab}$, solution of the vacuum equations.

\subsection{Antipodal transformations on the sphere}\label{AntipodalTransformations}
To link incoming and outgoing radiation at null infinity (see next subsection), we need to introduce the notion of antipodal points on the sphere. More specifically, the antipodal points on the sphere are those diametrically opposite to each other. Then, we define the antipodal transformation to be the one that carries a point in the sphere to its antipodal point.\\
In the usual spherical chart $(\theta,\phi)$, the antipodal transformation reads

\begin{align*}
    (\theta,\phi) \rightarrow (\pi-\theta,\pi+\phi) ,
\end{align*}

or in stereographic coordinates

\begin{align*}
    (\zeta,\bar{\zeta}) \rightarrow (-1/\bar{\zeta},-1/\zeta).
\end{align*}

We denote the antipodal transformation with the symbol $\widehat{}$, i.e, $\widehat{\zeta}=- 1/\bar{\zeta}$. 
In particular, if we write $l_-^a=\frac{1}{\sqrt{2}}(-1,r^i)$ with $r^i$ the corresponding spatial vector, the antipodal transformation is

\begin{align}\label{antipodall}
    \widehat{l}_-^a=\frac{1}{\sqrt{2}}(-1,\widehat{r}^i)=\frac{1}{\sqrt{2}}(-1,-r^i)=-\frac{1}{\sqrt{2}}(1,r^i)=-l_+^a.
\end{align}

Eq. (\ref{antipodall}) is the antipodal transformation applied to a null vector $l_-^a$ defined at past null infinity. Hence, this vector has its antipodal point at minus the position of a vector defined at future null infinity.\\
The antipodal transformation on the derivative operator, which will appear later in the calculations, can be proven to be
$$
\widehat{\eth}=-\bar{\eth}\ .
$$

\bigskip
\subsection{Scattering of gravitational waves}

If we impose the condition
\begin{equation}\label{metrics}
g^+_{ab}=g^-_{ab},
\end{equation}
then there is a correlation between the incoming and outgoing radiation, they are no longer free. As one is used to a causal picture, we can assume that we have free incoming gravitational radiation and want to know the effect on the outgoing radiation. A linear theory would immediately show that they are unscattered but Einstein equations, being highly non linear, should exhibit gravitational tails for the outgoing radiation.

At a linearized level eq. \ref{g2} yields,$$h^+_{1ab}=h^-_{1ab},$$ or $$l^{+a} \partial_{a}Z^+_{1}+l^{-a}\partial_{a}Z^-_{1}=0,$$ from which one obtains
\begin{equation}\label{PrimerordenNSF}
    \sigma^{+}_{1}(u,\zeta)+\overline{\sigma}^{-}_{1}(u,\hat{\zeta})=0,
\end{equation}
where $\hat{\zeta}$ is the antipodal point of $\zeta$ on the sphere. This relation will be used later to describe the trivial scattering of quantum gravitons at the linearized level, since both $\zeta$ and $\hat{\zeta}$ describe the same incoming and outgoing directions of a graviton with momentum $\vec{k}$. It is thus very useful to write down the solutions of the above defined fields using a Fourier decomposition of the incoming and outgoing fields.

To proceed further we assume $\sigma^{+}$ can be written as a small deviation of the outgoing graviton $\sigma^{+}_{1}$ :
\begin{equation}
    \sigma^{+}(u,\zeta)=\sigma^{+}_{1}(u,\zeta)+\sigma^{+}_{2}(u,\zeta),
\end{equation}
where $\sigma^{+}_{1}$ satisfies eq. \ref{PrimerordenNSF}. In this way, $\sigma^{+}_{2}$ yields the non trivial part of the scattering of classical gravitational waves or quantum gravitons.

We assume the outgoing Bondi Shear has a positive frequency decomposition 
\begin{equation}
    \sigma^{+}(u, \theta, \varphi) = \int_{0}^{\infty} \sigma^{+}(w, \theta, \varphi) e^{-iwu} \, dw
\end{equation}

Defining the Fourier transform of $\Omega^+_2$ as
\begin{align}\label{Omega2ka}
8\Omega^+_2(k^a,\zeta,\bar{\zeta}) &= \int d^4xe^{i x^ak_a}[\partial _{r}\Lambda^+_1 \partial _{r}\bar{\Lambda}^+_1+ \int _{r}^\infty dr' \int _{r'}^\infty dr''\partial _{r''}^{2}\Lambda^+_1 \partial _{r''}^{2}\bar{\Lambda}^+_1],
\end{align}
and following a calculation derived in the Appendix of ref.\cite{bordcoch2023asymptotic}, we obtain

\begin{align}\label{Omega2kb}
8\Omega^+_2(k^a,\zeta) &= \int \frac{d^3k_1}{2w_1}\frac{d^3k_2}{2w_2}\delta^4(k^a -(k_1-k_2)^a)\sigma^{+}(k_1)\bar{\sigma}^{+}(k_2)\mathcal{S}^+_{\Omega}(\zeta,k_1,k_2),
\end{align}
with
\begin{align}
\mathcal{S}^+_{\Omega}=G_{2,2'}(\zeta,\hat{k}_2)G_{-2,-2'}(\zeta,\hat{k}_1)\left( l^{+a}k_{1a}l^{+b}k_{2b}+ \frac{(l^{+a}k_{1a}l^{+b}k_{2b})^2}{(l^{+a}(k_1+k_2)_a)^2}\right),
\end{align}
and $l^{+a}=l^{+a}(\zeta)$, $G_{2,2'}(\zeta)=\eth^2\eth'^2G_{0,0'}$, $G_{-2,-2'}(\zeta)=\bar{\eth}^2\bar{\eth}'^2G_{0,0'}$.
Inverting \cref{Omega2kb} we obtain,
\begin{align}\label{Omega2x}
8\Omega^+_2(x^a,\zeta) &= \int \frac{d^3k_1}{2w_1}\frac{d^3k_2}{2w_2}e^{-i x^c (k_1-k_2)_c}\sigma^{+}(k_1)\bar{\sigma}^{+}(k_2)\mathcal{S}^+_{\Omega}(\zeta,k_1,k_2).
\end{align}
The second order solution can be split in two parts. One only involves an integral on the light cone cut and it si given by,
\begin{equation}\label{Z2cut}
    Z^+_{cut}=\oint d^2\hat{k}\left(G_{0,-2}(\zeta,\hat{k})[\sigma_1^+(Z_1,\hat{k})+ \sigma_2^+(Z^+_0,\hat{k})]+c.c.+G_{0,0}(\zeta,\hat{k})\Sigma^+(Z^+_0,\hat{k})\right),
\end{equation}
Note that in this equation $\sigma_1^+$ is completely determined by the incoming null data whereas $\sigma_2^+$ will be determined by the scattering process.

Likewise, one can get the non trivial part of $Z_2$ as the contribution from the integral on the future null cone from $x^a$, namely
\begin{align}\label{Z2cones}
Z^{+}_{2,cone}=- \oint d^2\hat{k} G_{0,0}(\zeta,\hat{k})\int_0^\infty ds   \big(2 \eth \bar{\eth}\Omega_2(y^c,\hat{k})+\eta^{ab}\partial_a \Lambda_1\partial_b \bar{\Lambda_1}(y^c,\hat{k})\big) ,
\end{align}
with $y^c= x^c+s l'^{c}$. Defining $N^+_x$ as the future null cone from the point $x^c$ and $C^+_x$ as the intersection of $N^+_x$ with future null infinity,  the 2-surface integral in \cref{Z2cut} is given on $C^+_x$ while the 3-dim integral is given on future null infinity and its boundary is $C^+_x$. The 3- dim integral of \cref{Z2cones} is given on $N^+_x$. An analogous formula can be given for $Z^{-}_{2,cone}$.

Following a calculation derived in the Appendix of ref.\cite{bordcoch2023asymptotic} one can show that

\begin{equation}\label{SegunoNSF}
    \sigma^{+}_2(u, \zeta, \overline{\zeta}) 
     = 4i \int 
    \frac{d^{3}k_{1}}{2w_{1}} 
    \frac{d^{3}k_{2}}{2w_{2}}
    \big(
    \begin{aligned}[t]
        &
        \sigma^{-}(\vec{k}_{1}) \overline{\sigma}^{-}(\vec{k}_{2}) e^{-iu|\vec{k}_{1} - \vec{k}_{2}|} 
        \left[S_{\Omega}(k_{1}, k_{2}, \zeta) + S_{A}(k_{1}, k_{2}, \zeta) \right] \\
        &- 
        \sigma^{-}(\vec{k}_{1}) \sigma^{-}(\vec{k}_{2}) 
        e^{iu|\vec{k}_{1} + \vec{k}_{2}|} 
        S_{B}(k_{1}, k_{2}, \zeta) \\
        &+ 
        \overline{\sigma}^{-}(\vec{k}_{1}) \overline{\sigma}^{-}(\vec{k}_{2}) 
        e^{-iu|\vec{k}_{1} + \vec{k}_{2}|} 
        \overline{S_{B}}(k_{1}, k_{2}, \zeta)\big),
    \end{aligned}
\end{equation}
with
\begin{align}
    S_{A}(k_{1},k_{2},\zeta)&=\dfrac{l^{+a}_1l^{+}_{a2}}{l^{+c}(k_{1}-k_{2})_{c}}\big(\delta^2(\zeta-\zeta_{1})\delta^2(\zeta-\zeta_{2})+G_{2,2'}(\zeta,\zeta_{1}) G_{-2,-2'}(\zeta,\zeta_{2})\big)\\
    S_{B}(k_{1},k_{2},\zeta)&=\dfrac{l^{+a}_1l^{+}_{a2}}{l^{+c}(k_{1}+k_{2})_{c}}\delta^2(\zeta-\zeta_{1}) G_{-2,-2'}(\zeta,\zeta_{2}),
\end{align}
where we have assumed the incoming radiation has compact support. It is important to note that the above result differs from the one obtained in \cite{bordcoch2023asymptotic} and gives the correct form of the nontrivial part of the scattering cross section. This is crucial at the quantum level, while the previous result yields a trivial scattering for quantum gravitons the above formula does not.

The corresponding Fourier transform of the above formula is given by,

\begin{equation}\label{2nd compacto}
    \sigma^{+}_2(w_k, \hat{k}) 
     = 4i \int 
    \frac{d^{3}k_{1}}{2w_{1}} 
    \frac{d^{3}k_{2}}{2w_{2}}
    \bigg(
    \begin{aligned}[t]
        &
        \sigma^{-}(\vec{k}_{1}) \overline{\sigma}^{-}(\vec{k}_{2})\delta(w-|\vec{k}_{1} - \vec{k}_{2}|) 
        \left[ S_{\Omega}(k_{1}, k_{2}, \vec{k}) + S_{A}(k_{1}, k_{2}, \vec{k}) \right] \\
        &- 
        \sigma^{-}(\vec{k}_{1}) \sigma^{-}(\vec{k}_{2}) 
        \delta(w-|\vec{k}_{1} + \vec{k}_{2}|) 
        S_{B}(k_{1}, k_{2}, \vec{k}) \\
        &+ 
        \overline{\sigma}^{-}(\vec{k}_{1}) \overline{\sigma}^{-}(\vec{k}_{2}) 
        \delta(w-|\vec{k}_{1} + \vec{k}_{2}|) 
        \overline{S_{B}}(k_{1}, k_{2}, \vec{k})\bigg),
    \end{aligned}
\end{equation}
with $(w_k, \hat{k})$ the components of the vector $\vec{k}$ in spherical coordinates.

\section{\label{sectionIII}Quantization of the null data}
Following Ashtekar's aymptotic quantization of the gravitational field\cite{ashtekar1987asymptotic,frittelli1997quantization}, we obtain commutation relations for the shear or the News tensor.\\

In General Relativity the shear $\sigma^{+}_{ab}$ or News tensor $N^{+}_{ab}$ at null infinity play the role of the initial field that will be promoted to a quantum operator.\\

In a given Bondi frame we have
\begin{align}
    \sigma^{+}_{ab}(u, \zeta) &= \overline{\sigma}{}^{+}(u, \zeta) m_{a} m_{b} + \sigma^{+}(u, \zeta) \overline{m}_{a} \overline{m}_{b},\label{3-11} \\
N^{+}_{ab} &= \dot{\sigma}{}^{+}_{ab}\\
    q_{ab} &= -m_{a}\overline{m}_{b}-\overline{m}_{a}m_{b}.\label{3-12}
\end{align}

Directly from the Poisson brackets \cite{ashtekar1987asymptotic,frittelli1997quantization}, promoting the fields to quantum operators, and using natural units with $c=\hbar=1$, one obtains\cite{frittelli1997quantization}

\begin{align}
    [\sigma^{+}(u, \zeta) , \overline{\sigma}{}^{+}(u', \zeta')] 
    &= -i \delta^2(\zeta - \zeta')  \Delta(u - u') \mathbb{I}, \label{3-22} \\
    [\overline{\sigma}{}^{+}(u, \zeta), \overline{\sigma}{}^{+}(u', \zeta')] 
    &= \sigma^{+}(u, \zeta), \sigma^{+}(u', \zeta')] = 0,  \label{3-23}
\end{align}
where $\Delta(u - u')$ is the sign function, a skew symmetric function with values $-\frac{1}{2}$ or $\frac{1}{2}$.
These commutation relations will be used throughout this work.
One can also obtain the commutation relations of the Fourier transform of $\sigma^{+}$. Starting with a positive frequency decomposition
\begin{equation}\label{positive decomp}
    \sigma^{+}(u, \theta, \varphi) = \int_{0}^{\infty} \sigma^{+}(w, \theta, \varphi) e^{-iwu} \,dw
\end{equation}

Directly from \ref{positive decomp} one gets:
\begin{equation}\label{3-conmut}
  [\sigma^{+}(w, \zeta), \overline{\sigma}^{+}(w', \zeta')] 
  = \frac{\delta(w - w')}{w} \delta^2(\zeta - \zeta').
\end{equation}

Since gravitons are massless fields they satisfy
\begin{equation}\label{3-esfera}
    k^{a}k_{a}=0 \implies w_{\vec{k}}=|\vec{k}|\implies  k^{a}=(w,w\vec{e}_{r})\quad \text{with $\vec{e}_{r}$ a unit vector on the sphere.}
\end{equation}
In our formulation the null vector in standard spherical coordinates can be written as $k^{a}=wl^{a}$, with $l^{a}l_{a}=0$, that has been used to define a null cone cut $Z_0$.

We also have $\sigma(\vec{k})=\sigma(w,\zeta)$.\\

Thus, formally the above relations can be written as
\begin{equation}\label{3-conmutacion}
    [\sigma^{+}(\vec{k}), \overline{\sigma}^{+}(\vec{k}')]= w_k\delta(\vec{k}-\vec{k}').
\end{equation}

Later on we will define the annihilation operator at future null infinity as $$a_{\mathrm{out}}(\vec{k}) = \sigma^+(w,\zeta),$$ 
 with the standard relation
\begin{equation}\label{3b-conmutacion}
    [a_{\mathrm{out}}(\vec{k}), a_{\mathrm{out}}^\dagger(\vec{k'})]=\omega_k \delta(\vec{k}-\vec{k}').\\
\end{equation}
A similar construction will be done at past null infinity.

\section{\label{sectionIV}Quantization of NSF}

Quantization of NSF cannot be achieved using the standard techniques of a Hamiltonian formulation, since two of the three field equations for $Z$ and $\Omega$ do not come from a Lagrangian. Nevertheless, as we saw in the previous section, from the Poisson brackets at null infinity one promotes the shear tensor and its complex conjugate at null infinity to quantum operators and they play the role of creation-annihilation operators. We thus adopt a field equation approach to quantize NSF, approach that can be seen in older textbooks (see for example ref. \cite{gasiorowicz} chapter 6). We thus assume that the field equations are also valid for the quantum variables of NSF, a standard assumption in quantum field theory, and that the free data obey the commutation relations found in the previous section. Here we explore the consequences of this quantization on the main variables of NSF, namely $Z$ and $\Omega$, since promoting free data to quantum operators automatically transforms $Z$ and $\Omega$ into quantum variables.

It is worth mentioning that once the NSF equations are written in a perturbative scheme with the points of the spacetime $x^a$ explicitly given, it becomes a "standard" quantum field theory. The operators $a_{out}(k)$ and $a_{in}(k)$ are eigenstates of energy and momentum and obey singular ccr. Therefore, one expects many divergences arising from the delta function of the ccr they satisfy, some of which are avoided by normal ordering.  One can regularize those operators as suitable wave packets when dealing with higher-order terms in the perturbation procedure, but as we will see below, singularities do not arise at a tree level. For higher terms one will certainly encounter divergences that must be addressed. One must also mention that coherent quantum states that are peaked at large initial data are excluded from this construction. 

Some results presented in this section have been obtained at a linear level before\cite{frittelli1997quantization}. However, the new point of view in this work is to obtain a perturbative quantum field theory in which the points of the spacetime are labels in the solution space of $Z$ and the metric is a quantum operator obtained from knowledge of $Z$ and $\Omega$. In this perturbation procedure to obtain a solution, the zeroth order term $Z_0$ is not quantized since it merely gives the four constants of integration $x^a$ associated with the solution. The points of space-time become labels on the quantum $Z$. Moreover, for fixed values of $(u,\zeta,\bar{\zeta})$ the hypersurface

\begin{align} \label{nullZ0}
Z_0 = x^a l_a = u = const.
\end{align}
is the past null cone (with respect to the flat metric) from the point $(u,\zeta,\bar{\zeta})$ at null infinity. From the point of view of the physical spacetime, it is a null plane with fixed direction $l_a$. The flat metric obtained from $Z$ is also a c-number in the quantized theory\\

As we showed in the previous section, the first order correction to the null flat foliation $Z_0$ is given by,

\begin{align}\label{Z1+}
     Z^{+}_1(x^a,\zeta,\bar{\zeta})= \oint_{S^2} \left(G_{0,-2'} \sigma^{+}(x^{a} l^{+\prime}_{a},\zeta',\bar{\zeta}')+G_{0,2'} \bar{\sigma}^{+}(x^{a} l^{+\prime}_{a},\zeta',\bar{\zeta}')\right) dS^{\prime},
 \end{align}
 and 
\begin{align}\label{metric1}
    h^+_{1ab}(x) =\frac{-1}{2\pi}\oint_{S^2} \left(m'_am'_b\dot{\bar{\sigma}}^{+}(x^c l^{+\prime}_c,\zeta',\bar{\zeta}')+ \; \; c.c. \right) dS^{\prime},
\end{align}
where we have used that $Z_0^{+}=u=x^a l^{+}_a$. Using a positive frequency decomposition (\ref{positive decomp}) for $\sigma^+$ and a spherical coordinate system for $k^a=\omega l^{+a}$ we  can rewrite the above equations as,

\begin{align}\label{Z1+b}
     Z^{+}_1(x^a,\zeta,\bar{\zeta})=  \int\dfrac{d^{3}k}{4\pi} \left(\frac{(l^a\overline{m}'_{a})^2}{l^bl'_b}\sigma^{+}(\vec{k})e^{-ik^{a}x_{a}}+\frac{(l^am'_a)^2}{l^bl'_b} \overline{\sigma}^{+}(\vec{k})e^{ik^{a}x_{a}} \; \; c.c. \right).
\end{align}

and

\begin{align}\label{FourierZ}
 h^{+}_{1ab}(x^{a})=\dfrac{-i}{\pi}\int \dfrac{d^{3}k}{2w_k}\left(\overline{m}_{a}\overline{m}_{b}\sigma^{+}(\vec{k})e^{-ik^{a}x_{a}}-m_{a}m_{b}\overline{\sigma}^{+}(\vec{k})e^{ik^{a}x_{a}}\right)   
\end{align}
with $m_a(k)$ orthogonal to $k_a$. We thus obtain a transverse spin 2 field on flat space-time where $\sigma^{+}(\vec{k})$ is an annihilation operator and the bivectors $m_am_b$ and c.c. span the 2 degrees of freedom of the quantum field.

Note that starting from a free field quantization of the shear at null infinity, and using NSF to construct the metric (\ref{metric1}) we end up with a standard spin 2 quantum field on Minkowski that has the standard form,
\begin{align}\label{FourierZ2}
 h^{+}_{ab}(x^{a})=\dfrac{-i}{\pi}\int \dfrac{d^{3}k}{2w_k}\left(\overline{m}_{a}\overline{m}_{b}a_{\mathrm{out}}(\vec{k})e^{-ik^{a}x_{a}}-m_{a}m_{b}a^\dagger_{\mathrm{out}}(\vec{k})e^{ik^{a}x_{a}}\right).   
\end{align}
It is important to remark that the Fourier transform of the shear, i.e., $\sigma^{+}(w,\zeta, \bar{\zeta})$, becomes the operator $a_{\mathrm{out}}(\vec{k})$ that is an eigenstate of the energy and momentum operators in a standard quantum field theory in flat space. This follows from the fact that $h^{+}_{ab}(x^{a})$ obeys the wave equation. Thus, one can write down a Lagrangian formulation where the energy, linear and angular momentum are conserved quantities.
It is also important to note that the quantization of the null data is given once and for all for any order of the perturbation procedure but the standard correlation with the Fourier decomposition given in eq \ref{FourierZ} will be lost, the second order perturbation metric does not satisfy the wave equation. Nevertheless, the definition of creation/anihilation for the Bondi shear at null infinity is used for the full perturbative approach since they are the null free data to construct empty asymptotically flat space times in the NSF, and we have promoted them to quantum operators.


\subsection{Trivial scattering of the linearized graviton solution}

 To obtain the scattering of incoming gravitons at past null infinity we first give the formula of the cut and metric using the retarded solutions, i.e,
 
 \begin{align}\label{Z1-b}
     Z^{-}_1(x^a,\zeta,\bar{\zeta})=  \frac{1}{4\pi}\oint_{S^2} \left(\frac{(l^{-a}m'_a)^2}{l^{-b}l'_{-b}} \bar{\sigma}^{-}(-x^cl'^{-}_c,\zeta',\bar{\zeta}')+ \; \; c.c. \right) dS^{\prime},
\end{align}
where we have used the null cone cut $Z^-_0=-x^a l^-_{a}$ ,  given by the intersection of the past null cone from $x^a$ with past null infinity and $m'_{+a}= m'_{-a}$. Note also that here $Z^-_0=v$, with $v$ the usual advanced time coordinate. Likewise, the metric can be written as,

\begin{align}\label{h1-}
    h^-_{1ab}(x)=\frac{1}{2\pi}\oint_{S^2} \left(m'_a m'_b \dot{\bar{\sigma}}^{-}(-x^c l'{}^{-}_c,\zeta',\bar{\zeta}')+ \; \; c.c. \right) dS^{\prime}.
\end{align}

Directly from
\begin{align}\label{h1b}
    h^{+}_{1ab}(x^c)=h^{-}_{1ab}(x^c),
\end{align}
one obtains the relationship between the free data at future and past null infinity by doing a coordinate transformation $\zeta \rightarrow \hat{\zeta}$ on  \cref{h1-} obtaining
\begin{align}\label{h1-hat}
    h^-_{1ab}(x)=\frac{1}{2\pi}\oint_{S^2} \left(m'_a m'_b \dot{\sigma}^{-}(x^a l^{+'}_a,\widehat{\zeta}',\widehat{\bar{\zeta}}')+ \; \; c.c. \right) dS^{\prime},
\end{align}
identifying the integrands in each expression for the linearized metric one obtains

\begin{align}\label{trivial}
&\dot{\sigma}^+(u, \zeta,\bar{\zeta})+\dot{\bar{\sigma}}^-(u, \widehat{\zeta},\widehat{\bar{\zeta}}')=0,
\end{align}
relating the incoming and outgoing gravitational radiation. The functional dependence of $\dot{\sigma}^+(u, \zeta,\bar{\zeta})$ on the antipodal value of $\dot{\bar{\sigma}}^-(u, \zeta,\bar{\zeta})$ follows directly from the fact that a null ray coming from a given direction in spherical coordinates goes to the antipodal direction after passing from the origin. At a quantum level on then has
\begin{align}\label{a trivial}
a_{\mathrm{out}}(\vec{k}) =\sigma^+(w, \zeta,\bar{\zeta})= -\bar{\sigma}^-(w, \widehat{\zeta},\widehat{\bar{\zeta}})=a_{\mathrm{in}}(\vec{k}),
\end{align}
giving a trivial scattering at the linear level. One should not be surprised that linear quantum gravitons behave like any free field that propagates along null geodesics on flat space.

Nevertheless, eq.(\ref{a trivial}) is telling us that the quantum operators defined from the free data at past and future null infinity are the incoming or outgoing free fields at the boundaries. We will retain this feature for non trivial scattering in the next section

\section{Non-trivial scattering of quantum gravitons}\label{sectionV}

The non-trivial scattering for gravitational waves can now be converted into a unitary transformation from the in and out states, i.e, we define the annihilation operators at future and past null infinity as,
\begin{align}\label{a non trivial}
a_{\mathrm{out}}(\vec{k}) &:=\sigma^+(w, \zeta,\bar{\zeta})\\
a_{\mathrm{in}}(\vec{k})&:=-\overline{\sigma}^-(w, \widehat{\zeta},\widehat{\bar{\zeta}}),
\end{align}
each of them satisfying the standard commutations relations that arise from the Poisson brackets at future and past null infinity,
\begin{equation}
[ a_{\mathrm{out}}(\vec{k}), a_{\mathrm{out}}^{\dagger}(\vec{k}')] =[ a_{\mathrm{in}}(\vec{k}), a_{\mathrm{in}}^{\dagger}(\vec{k}')] = \omega\delta(\vec{k}-\vec{k}'),
\end{equation}
and search for a unitary transformation of the form,
\begin{equation}\label{unitary}
    a_{\mathrm{out}}(\vec{k})=S^\dagger a_{\mathrm{in}}(\vec{k})S.
\end{equation}
with $$S=e^{i\epsilon\delta T}.$$
To first order (\ref{unitary}) yields

\begin{equation}
    a_{\mathrm{out}}(\vec{k})=a_{\mathrm{in}}(\vec{k})+i\epsilon [a_{\mathrm{in}},\delta T]=a_{\mathrm{in}}(\vec{k})+i\epsilon\delta a_{\mathrm{in}}(\vec{k}).
\end{equation}
For $\epsilon=0$ we obtain the trivial scattering \eqref{PrimerordenNSF}, whereas for $\epsilon=1$ we use the explicit form of the field equations to obtain 
\begin{equation}\label{4-51}
    \delta a_{\mathrm{in}}(\vec{k})= 4i \int 
    \frac{d^{3}k_{1}}{2w_{1}} 
    \frac{d^{3}k_{2}}{2w_{2}}
    \big(
    \begin{aligned}[t]
        &
        a^{\dagger}_{\mathrm{in}}(\vec{k}_{1}) a_{\mathrm{in}}(\vec{k}_{2}) \delta(w-|\vec{k}_{1} - \vec{k}_{2}|)
        \big[S_{\Omega}(k_{1}, k_{2}, \vec{k}) + S_{A}(k_{1}, k_{2}, \vec{k}) \big] \\
        &- 
        a^{\dagger}_{\mathrm{in}}(\vec{k}_{1}) a^{\dagger}_{\mathrm{in}}(\vec{k}_{2}) 
        \delta(w-|\vec{k}_{1} + \vec{k}_{2}|) 
        S_{B}(k_{1}, k_{2}, \vec{k}) \\
        &+ 
        a_{\mathrm{in}}(\vec{k}_{1}) a_{\mathrm{in}}(\vec{k}_{2}) 
        \delta(w-|\vec{k}_{1} + \vec{k}_{2}|) 
        \overline{S_{B}}(k_{1}, k_{2}, \vec{k})\big),
    \end{aligned}
\end{equation}
The above equations can be used to obtain the S-matrix coefficients for any scattering process assuming the free field operators obey the standard relations.

Assuming that we have a vacuum state $|0\rangle$ that satisfies 

\begin{equation}
        a(\vec{k})|0\rangle=0
    \end{equation}

    and that the one graviton state with momentum $\vec{k}$ is given by
    \begin{equation}
        a^{\dagger}(\vec{k})|0\rangle=|1\rangle_{\vec{k}},
    \end{equation}
     etc. we can compute the scattering process of any number of incoming or outoging gravitons.

   As an example we compute the scattering of a 2 particle state. Given the incoming state $a_{\mathrm{in}}^{\dagger}(\vec{k}_{1}) a_{\mathrm{in}}^{\dagger}(\vec{k}_{2}) | 0 \rangle $ , and outgoing state $a_{\mathrm{out}}^{\dagger}(\vec{k}'_{1}) a_{\mathrm{out}}^{\dagger}(\vec{k}'_{2}) | 0 \rangle $ , the scattering matrix coefficient is given by
    \begin{equation}\label{3-33}
        S_{k'_{1}k'_{2} k_{1}k_{2}}= \langle k'_{1}k'_{2} |k_{1}k_{2}  \rangle= \langle 0  0 | a_{\mathrm{out}}(\vec{k}'_{1})a_{\mathrm{out}}(\vec{k}'_{2})a_{\mathrm{in}}^{\dagger}(\vec{k}_{2})a_{\mathrm{in}}^{\dagger}(\vec{k}_{1}) | 0  0 \rangle,  
    \end{equation}
    and the transition probability  $k_{1},k_{2}\rightarrow k'_{1},k'_{2}$ given by
    \begin{equation}
        \mathcal{P}_{k_{1},k_{2}\rightarrow k'_{1},k'_{2}}=|\langle k'_{1}k'_{2} |k_{1}k_{2}  \rangle|^{2}=|S_{k'_{1}k'_{2} k_{1}k_{2}}|^{2}
    \end{equation}

Using eq. (\eqref{4-51}) we obtain,

\begin{align}
    S_{k'_{1}k'_{2} k_{1}k_{2}} &= \langle 0  0 | a_{\mathrm{out}}(\vec{k}'_{1})a_{\mathrm{out}}(\vec{k}'_{2})a_{\mathrm{in}}^{\dagger}(\vec{k}_{2})a_{\mathrm{in}}^{\dagger}(\vec{k}_{1}) | 0  0 \rangle  \\
    &=\langle 0  0 |  
    \left( a_{\mathrm{in}}(\vec{k}'_{1}) + i \delta a_{\mathrm{in}}(\vec{k}'_{1}) \right)  
    \left( a_{\mathrm{in}}(\vec{k}'_{2}) + i \delta a_{\mathrm{in}}(\vec{k}'_{2}) \right)  
    a_{\mathrm{in}}^{\dagger}(\vec{k}_{2}) a_{\mathrm{in}}^{\dagger}(\vec{k}_{1}) | 0  0 \rangle  \\
    &=   
    \lvert \vec{k}_{1} \rvert \delta^{3}(\vec{k}_{1}-\vec{k}'_{1})  
    \lvert \vec{k}_{2} \rvert \delta^{3}(\vec{k}_{2}-\vec{k}'_{2})  
    + i I_{1} + i I_{2} - I_{3}.
\end{align}

with 

\begin{align}
    I_{1}&=\langle 0  0 |a_{\mathrm{in}}(\vec{k}'_{1})\delta a_{\mathrm{in}}(\vec{k}'_{2})a_{\mathrm{in}}^{\dagger}(\vec{k}_{2})a_{\mathrm{in}}^{\dagger}(\vec{k}_{1})| 0  0 \rangle\label{Iuno}\\
    I_{2}&=\langle 0  0 |\delta a_{\mathrm{in}}(\vec{k}'_{1})a_{\mathrm{in}}(\vec{k}'_{2})a_{\mathrm{in}}^{\dagger}(\vec{k}_{2})a_{\mathrm{in}}^{\dagger}(\vec{k}_{1})| 0  0 \rangle\label{Idos}\\
    I_{3}&=\langle 0  0 |\delta a_{\mathrm{in}}(\vec{k}'_{1})\delta a_{\mathrm{in}}(\vec{k}'_{2})a_{\mathrm{in}}^{\dagger}(\vec{k}_{2})a_{\mathrm{in}}^{\dagger}(\vec{k}_{1})| 0  0 \rangle\label{Itres}
\end{align}

The only non vanishing term for this process is
\begin{equation}\label{I3}
    I_{3} = 
    \int \frac{d^{3}k_{3}}{2w_{3}} \frac{d^{3}k_{4}}{2w_{4}} 
         \frac{d^{3}k_{5}}{2w_{5}} \frac{d^{3}k_{6}}{2w_{6}} 
    \begin{aligned}[t]
        &\tau_{B}(\vec{k}'_{1}, k_{3}, k_{4}) \, \tau_{\chi}(\vec{k}'_{2}, k_{5}, k_{6}) \\
        &\times \langle 0 0 | a_{\mathrm{in}}(\vec{k}_{3}) a_{\mathrm{in}}(\vec{k}_{4}) 
        a^{\dagger}_{\mathrm{in}}(\vec{k}_{5}) a_{\mathrm{in}}(\vec{k}_{6}) 
        a_{\mathrm{in}}^{\dagger}(\vec{k}_{2}) a_{\mathrm{in}}^{\dagger}(\vec{k}_{1}) | 0 0 \rangle
    \end{aligned}
\end{equation}
with
\begin{align}
    \tau_{\chi}(\vec{k}'_{2}, \vec{k}_{5}, \vec{k}_{6}) &= 
    \begin{aligned}[t]
        &\,\delta(w'_{2} - |\vec{k}_{5} - \vec{k}_{6}|) \big(S_{\Omega}(\vec{k}_{5}, \vec{k}_{6}, \vec{k}'_{2}) + 
    S_{A}(\vec{k}_{5}, \vec{k}_{6}, \vec{k}'_{2})\big),
    \end{aligned} \\
    \tau_{B}(\vec{k}'_{1}, \vec{k}_{3}, \vec{k}_{4}) &= 
    \delta(w'_{1} - |\vec{k}_{3} + \vec{k}_{4}|) \, 
    \overline{S_{B}}(\vec{k}_{3}, \vec{k}_{4}, \vec{k}'_{1}).
\end{align}

A long but straightforward calculation gives,

\begin{equation}
    S_{k'_{1}k'_{2} k_{1}k_{2}} = 
    \int \frac{d^{3}k}{2w_k}\Big(\tau_{B}(\vec{k}'_{1}, \vec{k}_{1}, \vec{k})  \tau_{\chi}(\vec{k}'_{2},\vec{k}, \vec{k}_{2})+\tau_{B}(\vec{k}'_{1}, \vec{k}_{2}, \vec{k})  \tau_{\chi}(\vec{k}'_{2}, \vec{k}, \vec{k}_{1})\Big).
\end{equation}
The first term can be called the direct term whereas the second one as the exchange term if one makes an analogy with electron-electron Moller scattering in QED.

\section{Summary and conclusions}\label{sectionVI} \nocite{*}

In this work we have used the Null Surface Formulation of General Relativity to quantize
asymptotically flat spaces times that are a small perturbation of Minkowski space\cite{Chrusciel2002}, and for small initial null data that yields regular closed cuts both at future and past null infinity. 

Using the available vacuum field equation for the main variables of the formalism
we first present the solutions by a perturbation scheme valid up to 2nd order and
obtain advanced and retarded solutions that depend on the future or past radiation
data representing outgoing or incoming gravitational waves.
We then obtain a quantum scattering theory for the asymptotic quantization
procedure by promoting the null free data at future or past null infinity to quantum
operators.[5]. In our formalism the same free data is used for any order of the
perturbation procedure, i.e., the phase space is constructed once and for all in our
approach. This is extremely important at a classical or quantum level since one can
introduce either a canonical form with Poisson brackets or quantum commutation
relations for fields given on the null boundaries that will not be modified as one proceeds
with a perturbation calculation. The calculations performed at second order do not
have divergences if one assumes peeling for the gravitational field. Thus, a future area
of research is to derive ”Feynman-like” formulae for the higher order terms and analyze
possible divergences, compute radiative corrections, etc.

One drawback of this approach is  that coherent states peaked around large initial Bondi data are ruled out in this construction. This follows directly from the assumption of a weak Bondi data. It is worth pursuing an alternative approach to include those states which are part of the Hilbert space on the asympotitc quantization.

It is also worth comparing this approach with other perturbed formulations such as the covariant quantization of gravity since both are written as a perturbative scheme. 

The two approaches are quite different. In NSF, null infinity is given once and for all together with the free data, the News function. At every order of the perturbation procedure, one keeps modifying the null cuts, or equivalently, the conformal structure of the space-time so that the non trivial cuts are consistent with the free data. Thus, it is natural to quantize the free data using Ashetkar's free field quantization at null infinity. What NSF shows is that these quantum operators become creation/annihilation operators. At a linearized level they are are eigenstates of energy and momentum of linearized gravity, thus giving similar results than the covariant approach. However, the second order calculation is radically different. In NSF the non trivial scattering term uses the modified null cone cut to obtain the results presented in this work whereas the covariant quantization keeps using the flat characteristic surfaces throughout the perturbation calculation. Thus, it is interesting to see which approach appears to be more reasonable.

One should mention that even the starting classical field equations of covariant quantization fail to produce an acceptable perturbative theory for the following: the classical equations use flat characteristic surfaces in the whole perturbation scheme. For example, they use the flat Dalambertian operator and its Green function at every step of the perturbation procedure. The hidden problem in this procedure is that the flat null cones do not reach null infinity for the non-flat metric they are trying to obtain. These surfaces end either at time-like or space-like infinity and fail to make contact with null infinity.
From then on it is impossible to compare the quantum in and out states that are given in NSF with the similar objects in the covariant approach in the sense that one does not know where are the in and out states defined in the covariant approach, certainly not at null infinity.

There is also an assumption that has been used in this work, namely, the existence
of a map between the past and future Bondi coordinates mediated by a regularity
assumption of spacelike infinity[7, 9]. This assumption is true for the compactified flat
space-time that has been used in this work. However, if one does not have such a map
then the scattering matrix construction would have an S2 gauge freedom. Although
some progress has been made to understand the regularity structure of spacelike infinity
on a generic situation, there is still much work to be done to get a full understanding
of its conformal completion[17, 18].

It is also important to remark that the quantization of NSF in 2+1 dimensions coupled to an axisymmetric scalar field gives very interesting results\cite{Dominguez-Tiglio,Ashtekar2022}. There, one finds that space-time points in the interior become
fuzzy in a specific, well-defined sense but the quantum fluctuations
of the cut function $Z$ at infinity are highly suppressed. It should be very interesting to compare these results with a quantization of NSF coupling the gravitational field with a scalar field in 4 dimensions\cite{Kozameh-Tassone}. 

\vspace{60pt}
\bibliographystyle{iopart-num}
\bibliography{bibliography}
\end{document}